# Unusually large enhancement of thermopower in an electric field induced two-dimensional electron gas

*By* Hiromichi Ohta\*, Taku Mizuno, Shijian Zheng, Takeharu Kato, Yuichi Ikuhara, Katsumi Abe, Hideya Kumomi, Kenji Nomura, *and* Hideo Hosono

Prof. Hiromichi Ohta
Graduate School of Engineering, Nagoya University, Furo-cho, Chikusa, Nagoya 464–8603, Japan / PRESTO, Japan Science and Technology Agency, 5 Sanbancho, Chiyoda, Tokyo 102–0075 (Japan)

Mr. Taku Mizuno
Graduate School of Engineering, Nagoya University, Furo-cho, Chikusa, Nagoya 464–8603 (Japan)

Dr. Shijian Zheng, and Dr. Takeharu Kato
Nanostructures Research Laboratory, Japan Fine Ceramics Center, 2–4–1 Mutsuno, Atsuta, Nagoya 456–8587 (Japan)

Prof. Yuichi Ikuhara
Nanostructures Research Laboratory, Japan Fine Ceramics Center, 2–4–1 Mutsuno, Atsuta, Nagoya 456–8587, Japan / Institute of Engineering Innovation, The University of Tokyo, 2–11–16 Yayoi, Bunkyo, Tokyo 113–8656 (Japan)

Mr. Katsumi Abe
Canon Inc., 3–30–2 Shimomaruko, Ohta-ku, Tokyo 146–8501, Japan / Frontier Research Center, Tokyo Institute of Technology, 4259 Nagatsuta, Midori, Yokohama 226–8503 (Japan)

Dr. Hideya Kumomi
Canon Inc., 3–30–2 Shimomaruko, Ohta-ku, Tokyo 146–8501 (Japan)

Prof. Kenji Nomura, and Prof. Hideo Hosono
Frontier Research Center, Tokyo Institute of Technology, 4259 Nagatsuta, Midori, Yokohama 226–8503 (Japan)

Thermoelectric energy conversion technology attracts attention to convert waste heats from various sources, *e.g.*, electric power plants, factories, automobiles, computers, and even human bodies, into electricity.[1−3] In 1821, T. J. Seebeck discovered the principle of thermoelectric energy conversion: a thermo-electromotive force ($\Delta V$) is generated between two ends of a metal bar by introducing a temperature difference ($\Delta T$) in the bar.[4] The value of $\Delta V/\Delta T$ is so-called thermopower or Seebeck coefficient ($S$), which is an important physical parameter to obtain high thermoelectric figure of merit, $ZT = S^2 \cdot \sigma \cdot T \cdot \kappa^{-1}$, where $Z$, $T$, $\sigma$ and $\kappa$ are a figure of merit, the absolute temperature, the electrical conductivity, and the thermal conductivity, respectively.

The |$S$| value of a thermoelectric material can be expressed by the following Mott equation [5]

$$S = \frac{\pi^2}{3} \frac{k_B^2 T}{e} \left\{ \frac{d[\ln(\sigma(E))]}{dE} \right\}_{E=E_F}$$

$$= \frac{\pi^2}{3} \frac{k_B^2 T}{e} \left\{ \frac{1}{n} \cdot \frac{dn(E)}{dE} + \frac{1}{\mu} \cdot \frac{d\mu(E)}{dE} \right\}_{E=E_F}$$

where $k_B$, $e$, $n$, and $\mu$ are the Boltzmann constant, electron charge, carrier concentration, and carrier mobility, respectively. The $S^2 \cdot \sigma$ value of a thermoelectric material must be maximized by means of $n$ because of the commonly observed trade-off relationship between $S$ and $\sigma$ in terms of $n$: $\sigma$ increases almost linearly with increasing $n$, while |$S$| decreases with $n$. In addition, |$S$| strongly depends on the energy derivative of the electronic density of states (DOS) at around the Fermi energy ($E_F$), $\left[ \frac{\partial DOS(E)}{\partial E} \right]_{E=E_F}$. Thus, the $ZT$ value of thermoelectric materials can be dramatically enhanced by modifying the DOS in low-dimensional structures such as two-dimensional quantum-wells, [6−8] due to that an enhancement of |$S$| occurs. Therefore, two-dimensionally confined electrons showing

unusually large |S| have attracted attention as a potential approach for developing high performance thermoelectric materials.

Here we hypothesize that unusually large enhancement of |S| should be observed in an electric field induced two-dimensional electron gas (2DEG). **Figure 1a** schematically depicts a conventional metal-insulator-semiconductor (n-type) field effect transistor structure on a thermoelectric material. For a positive gate voltage ($V_g$), a 2DEG is formed at the gate insulator/thermoelectric material interface. The sheet carrier concentration ($n_{sheet}$) of the 2DEG increases monotonically with the applied electric field, while the thickness ($t_{eff}$) becomes extremely thin (**Fig. 1b**) because the residual electrons in a thermoelectric material are strongly attracted by the electric field, leading to an ultimate modification of the DOS. Because the |S| value decreases as $n_{sheet}$ increases, the |S| value of the 2DEG also decreases with the applied electric field in the bulk state. However, we hypothesize that unusually large enhancement of |S| can be observed when the $t_{eff}$ is thinner than the critical thickness for DOS modification. Recently, electric field modulations of |S| for rubrene, [9] PbSe nanowires, [10] and SrTiO$_3$ [11] have been demonstrated. However, unusually large enhancement of |S| has not been observed: the |S| value decreased monotonically with electric field due to an increase in the $n$ value, most likely due to that the 2DEG was thicker than the critical thickness.

To verify our hypothesis, we explored an excellent gate insulator, which can strongly accumulate electrons into an extremely thin 2DEG layer. Although liquid electrolytes including "gel" would be very useful to accumulate electrons at the 2DEG using their huge capacitance, [12–14] they would not be suitable for the present application without sealing due to liquid leakage problem. Very recently, we discovered that water-infiltrated nanoporous 12CaO·7Al$_2$O$_3$ glass (Calcium Aluminate with Nanopores, CAN, hereafter) [15] works as an excellent gate insulator, and it can strongly accumulate electrons up to $n_{sheet}$ ~10$^{15}$ cm$^{-2}$, which is two orders of magnitude higher than that accumulated by conventional gate insulators ($n_{sheet}$

~$10^{13}$ cm$^{-2}$). It should be noted that CAN is a chemically stable rigid glassy solid, showing excellent adhesion with oxide surface and no water leakage.

Here we demonstrate that an electric field induced 2DEG yields unusually large enhancement of |$S$|. Using CAN as the gate insulator, we fabricated a field effect transistor (FET) structure (**Fig. 2a**) on a SrTiO$_3$ crystal, which is a promising candidate as a thermoelectric material.[16, 17] An electric field application to the CAN-gated SrTiO$_3$ FET provides an extremely thin (~2 nm) 2DEG layer with $n_{sheet}$ ~2×10$^{15}$ cm$^{-2}$. As a result, unusually large enhancement of |$S$| for the 2DEG, approximately five times larger than that of the bulk, was clearly observed.

**Figure 2b** shows typical transfer characteristic curve of the CAN-gated SrTiO$_3$ FET at room temperature (RT). A very large anticlockwise hysteresis of drain current ($I_d$) with a rather large gate current ($I_g$) was observed, indicating motion of the H$^+$ (H$_3$O$^+$) and OH$^-$ ions in the CAN film. Detail of the transistor characteristics is described elsewhere. [15] It should be noted that water electrolysis occurs in the CAN film during the $V_g$ application. Before applying $V_g$ (virgin state), trilayer structure composed of Ti/CAN (~200 nm)/SrTiO$_3$ was clearly observed in the cross-sectional TEM image (**Fig. 2c**). Large amount of light spots (diameter ~10 nm), indicating infiltrated water, are seen in the whole CAN region. However, after positive $V_g$ (+40 V) application (**Fig. 2d**), thickness of the CAN layer dramatically increased (~550 nm), ~2.8 times thicker than the virgin state. Further, rather large light spots (50–200 nm) are seen in the whole CAN region, clearly indicating that water electrolysis (generation of H$_2$ and O$_2$ gases) occurred during the positive $V_g$ application. These results suggest that, for positive $V_g$, H$^+$ (H$_3$O$^+$) ions in the CAN film strongly attract residual electrons in the SrTiO$_3$ crystal, leading to the formation of 2DEG. After that, water electrolysis occurs at the SrTiO$_3$ and the Ti surfaces.

Here we show that an electric field induced 2DEG yields unusually large enhancement of |$S$|. **Figures 3** (left panels) show the changes of (a) |$S$| and (b) $n_{sheet}$ for the 2DEG in the

CAN-gated SrTiO$_3$ FET (red) as a function of $V_g$ at RT. The Hall mobility values ($\mu_{Hall}$) were 2.4±0.6 cm$^2$ V$^{-1}$ s$^{-1}$, did not depend on the $n_{sheet}$. For comparison, the values for a dry C12A7-gated SrTiO$_3$-FET [11] (blue) are also plotted. When $V_g$ was smaller than +22 V, the |S| value for the 2DEG in the CAN-gated FET gradually decreased with $V_g$ due to an increase in $n_{sheet}$ from 1×10$^{13}$ to 2.5×10$^{14}$ cm$^{-2}$. However, when $V_g$ exceeded +22 V, the |S| value increased while $n_{sheet}$ continuously increased from 2.5×10$^{14}$ to 2×10$^{15}$ cm$^{-2}$ with $V_g$. This V-shaped turnaround (indicated by the arrow) of the |S|– $V_g$ behavior is very unusual. *cf.* |S| value monotonously decreased from 1,140 to 620 µV K$^{-1}$ for the 2DEG in the dry C12A7-gated FET due to a monotonic increase in $n_{sheet}$ from 6×10$^{11}$ to 2×10$^{13}$ cm$^{-2}$. It should be noted that $n_{sheet}$ of the 2DEG for the CAN-gated FET at $V_g$=40 V was ~2×10$^{15}$ cm$^{-2}$, which is two orders of magnitude higher than that for the dry C12A7-gated FET at the same $V_g$ (≈Electric field 2 MV cm$^{-1}$), clearly demonstrating the high electron accumulation ability of the CAN gate insulator.

Afterward we measured the retention time dependence of |S| and $n_{sheet}$ for the CAN-gated FET. **Figure 3** (right panel) shows the changes of (c) |S| and (d) $n_{sheet}$ for the 2DEG in the CAN-gated SrTiO$_3$-FET as a function of the $V_g$ retention time at RT. $V_g$ was varied as follows: +15 V to –15 V, and then to –20 V, –30 V and finally to –40 V. A V-shaped turnaround of |S| was also observed in the |S|–time relation (indicated by the arrow). For a positive $V_g$ (+15 V), |S| initially decreased from 700 to 250 µV K$^{-1}$ with time (up to 500 s, indicated by arrow), but then dramatically recovered to ~600 µV K$^{-1}$. However, the value of $n_{sheet}$ increased monotonically with time. The $n_{sheet}$ value at the minimum |S| was 2.5×10$^{14}$ cm$^{-2}$, which agreed well with the values in the |S|– $V_g$ relation (Figs. 3a, 3b). For a negative $V_g$, |S| gradually increased from ~600 to ~1,000 µV K$^{-1}$, while $n_{sheet}$ decreased slightly from 2×10$^{15}$ to 8×10$^{14}$ cm$^{-2}$.

To verify the observed |S| values for the 2DEG in the CAN-gated SrTiO$_3$ FET, the |S| values were plotted as a function of $n_{sheet}$ (**Fig. 4a**) and $t_{eff}$ (**Fig. 4b**) on a logarithmic scale.

Additionally, the simulated bulk |$S$| values are plotted (grey dotted line). Detailed information about the 2DEG simulation ($n_{sheet}$, $t_{eff}$, |$S$|) is described in **Figs. S1**, **S2** and **Table I**. It should be noted that the |$S$| values for the 2DEG in both the CAN- and the dry C12A7-gated FETs were smoothly connected. |$S$| initially decreased with $n_{sheet}$ from ~1,150 to ~250 µV K$^{-1}$. Simultaneously, the $t_{eff}$ decreased from ~200 to ~2 nm. In this region, the $n_{sheet}$ dependence of |$S$| was similar to that of the simulated bulk values. When $n_{sheet}$ exceeded ~2.5×10$^{14}$ cm$^{-2}$, |$S$| increased drastically to ~950 µV K$^{-1}$, while $t_{eff}$ remained nearly constant (~2 nm), demonstrating an unusually large |$S$| for the electric field induced 2DEG. The |$S$| of the 2DEG in the CAN-gated FET was reversibly modulated from 600 ($n_{sheet}$ ~2×10$^{15}$ cm$^{-2}$) to 950 µV K$^{-1}$ ($n_{sheet}$ ~8×10$^{14}$ cm$^{-2}$). The |$S$| vs. log $n_{sheet}$ relation was approximately five times larger than that that of the bulk. Thus, we successfully demonstrated that an electric field induced 2DEG yields unusually large enhancement of |$S$|.

The significance of the present results is that they demonstrate an electric field induced 2DEG can exhibit an unusually large |$S$|, which is approximately five times larger than that that of the bulk. Using water-infiltrated nanoporous 12CaO·7Al$_2$O$_3$ glass as the gate insulator, carrier electrons up to ~10$^{15}$ cm$^{-2}$ can accumulate within an extremely narrow 2DEG (~2 nm). Although artificial superlattice materials also exhibit an enhanced |$S$|, [6−8] fabrication of artificial superlattice structures using state-of-the-art thermoelectric materials [18−21] is extremely difficult due to their complicated crystal structures. Furthermore, the production cost of such superlattice materials, which can be fabricated by precise vacuum deposition methods such as molecular beam epitaxy (MBE) and pulsed laser deposition (PLD) at high temperatures, is extremely high. On the other hand, the present method only requires "CAN", which can be fabricated at RT with low energy requirements. Thus, the present electric field effect approach should be applicable to fully verify the performance of thermoelectric materials with complicated crystal structures. Additionally, this approach may accelerate the development of nanostructures of high performance thermoelectric materials.

In conclusion, we have demonstrated that an electric field induced 2DEG provides unusually large enhancement of |$S$|. We fabricated FETs on a SrTiO$_3$ plate using water-infiltrated nanoporous glass as a gate insulator, which can accumulate carrier electrons up to ~10$^{15}$ cm$^{-2}$. An electric field application provided an extremely thin (~2 nm) 2DEG, which exhibited unusually large |$S$| value with high $n_{sheet}$ up to ~2×10$^{15}$ cm$^{-2}$. The |$S$| for the 2DEG was modulated from ~600 ($n_{sheet}$ ~2×10$^{15}$ cm$^{-2}$) to ~950 μV K$^{-1}$ ($n_{sheet}$ ~8×10$^{14}$ cm$^{-2}$), and the |$S$| vs. log $n_{sheet}$ relation was approximately five times larger than that of the bulk, clearly demonstrating that an electric field induced 2DEG provides unusually large enhancement of |$S$|

Moreover, because the present electric field induced 2DEG approach is simple and effectively verifies the performance of thermoelectric materials, it may accelerate the development of nanostructures for high performance thermoelectric materials.

*Experimental*

*Device fabrication*: The FETs (**Fig. 2a**) were fabricated on the (001)-face of SrTiO$_3$ single crystal plates (10×10×0.5 mm$^3$, SHINKOSHA Co.), which were treated in a NH$_4$F-buffered HF (BHF) solution [22]. First, 20-nm-thick metallic Ti films, which served as the source (S) and drain (D) electrodes, were deposited through a stencil mask by electron beam (EB) evaporation (base pressure ~10$^{-4}$ Pa, without substrate heating/cooling) onto the SrTiO$_3$ plate. The channel length and width were 800 and 400 μm, respectively. Then a 200-nm-thick CAN film was deposited through a stencil mask by PLD (KrF excimer laser, fluence ~3 J cm$^{-2}$ pulse$^{-1}$) at RT using dense polycrystalline C12A7 ceramic as a target. During CAN deposition, the oxygen pressure in the deposition chamber was kept at 5 Pa. The bulk density of the resultant CAN film was ~2.07 g cm$^{-3}$, evaluated by grazing incidence X-ray reflectivity (data not shown), which corresponds to 71% of fully dense amorphous C12A7 (2.92 g cm$^{-3}$). AC conductivity of the CAN film was 2.2×10$^{-9}$ S cm$^{-1}$ at RT, which was ~4% of that of ultra pure water (5.5×10$^{-8}$ S cm$^{-1}$). Finally, a 20-nm-thick metallic Ti film, which was used as the gate electrode, was deposited through a stencil mask by EB evaporation. Additionally, four electrodes E1–E4 (Ti/Au) were used for the Hall voltage measurements.

*Thermopower measurements*: We measured $|S|$ and $n_{sheet}$ at RT during several $V_g$ applications. Because the FET exhibited a non-volatile memory-like behavior as shown in Fig. 2b, the properties were measured successively after applying $V_g$. For the $|S|$ measurements, we used two Peltier devices (Fig. 2a, denoted as cooling and heating), which were placed under the FET, to give a temperature difference between the S and D electrodes. Two thermocouples (K-type, 150 μm in diameter, SHINNETSU Co.), which were mechanically attached at both edges of the channel, monitored the temperature difference ($\Delta T$, 0–5 K). The thermo-electromotive force ($\Delta V$) and $\Delta T$ values were simultaneously measured at RT. The $S$-values were obtained from the slope of the $\Delta V$–$\Delta T$ plots. $n_{sheet}$ values were measured by the conventional dc four-probe method with the van der Pauw electrode configuration at RT.

*Acknowledgements*

We thank S-W. Kim and R. Asahi for the valuable discussions. H.O. is supported by MEXT (22360271, 22015009). The Research at Tokyo Tech. is supported by JSPS–FIRST Program. Supporting Information is available online from Wiley InterScience.

[1] D. M. Rowe, Ed. *CRC Handbook of Thermoelectrics* (CRC Press, Boca Raton, FL,**1995**).

[2] F. J. DiSalvo, *Science* **1999**, 285, 703.

[3] G. J. Snyder, E. S. Toberer, *Nature Mater.* **2008**, 7, 105.

[4] T. J. Seebeck, *Abh. K. Akad. Wiss*. **1823**, 265.

[5] M. Cutler, N. F. Mott, *Phys. Rev.* **1969**, 181, 1336.

[6] L. D. Hicks, M. S. Dresselhaus, *Phys. Rev. B* **1993**, 47, 12727.

[7] L. D. Hicks, T. C. Harman, X. Sun, M. S. Dresselhaus, *Phys. Rev. B* **1996**, 53, R10493.

[8] H. Ohta, S-W. Kim, Y. Mune, T. Mizoguchi, K. Nomura, S. Ohta, T. Nomura, Y.Nakanishi, Y. Ikuhara, M. Hirano, H. Hosono, K. Koumoto, *Nature Mater*. **2007**, 6, 129.

[9] K. P. Pernstich, B. Rössner, B. Batlogg, *Nature Mater*. **2008**, 7, 321.

[10] W. Liang, A. I. Hochbaum, M. Fardy, O. Rabin, M. Zhang, P. Yang, *Nano Lett*. **2009**, 9, 1689.

[11] H. Ohta, Y. Masuoka, R. Asahi, T. Kato, Y. Ikuhara, K. Nomura, H. Hosono, *Appl. Phys. Lett*. **2009**, 95, 113505.

[12] K. Ueno, S. Nakamura, H. Shimotani, A. Ohtomo, N. Kimura, T. Nojima, H. Aoki, Y. Iwasa, M. Kawasaki, *Nature Mater*. **2008**, 7, 855.

[13] L. Kergoat, L. Herlogsson, D, Braga, B. Piro, M. Pham, X. Crispin, M. Berggren, G. Horowitz, *Adv. Mater*. **2010**, 22, 2565.

[14] Y. Yamada, K. Ueno, T. Fukumura, H. T. Yuan, H. Shimotani, Y. Iwasa, L. Gu, S. Tsukimoto, Y. Ikuhara, M. Kawasaki, *Science* **2011**, 332, 1065.

[15] H. Ohta, Y. Sato, T. Kato, S-W. Kim, K. Nomura, Y. Ikuhara, H. Hosono, *Nature Commun*. **2010**, 1, 118.

[16] T. Okuda, K. Nakanishi, S. Miyasaka, Y. Tokura, *Phys. Rev. B* **2001**, 63, 113104.

[17] H. Ohta, *Mater. Today* **2007**, 10, 44.

[18] G. S. Nolas, J. L. Cohn, G. A. Slack, *Appl. Phys. Lett*. **1998**, 73, 178.

[19] S. J. Poon, *Semicond. Semimet*. **2001**, 70, 37.

[20] G. J. Snyder, M. Christensen, E. Nishibori, T. Caillat, B. Iversen, *Nature Mater*. **2004**, 3, 458.

[21] J-S. Rhyee, K-H. Lee, S-M. Lee, E. Cho, S-I. Kim, E. Lee, Y-S. Kwon, J-H. Shim, G. Kotliar, *Nature* **2009**, 459, 965.

[22] M. Kawasaki, K. Takahashi, T. Maeda, R. Tsuchiya, M. Shinohara, O. Ishiyama, T. Yonezawa, M. Yoshimoto, H. Koinuma, *Science* **1994**, *266*, 1540.

**Figure 1.** Unusually large enhancement of thermopower in an electric field induced two-dimensional electron gas. (a) Schematic illustration of a conventional metal-insulator-semiconductor (n-type) field effect transistor structure on a thermoelectric material. For a positive $V_g$, a 2DEG layer is formed at the gate insulator/semiconductor interface. $S$ of the 2DEG layer (effective thickness: $t_{eff}$, sheet carrier concentration: $n_{sheet}$) is measured by introducing a temperature gradient at both ends of the 2DEG layer. (b) Electric field modulation of $|S|$, $n_{sheet}$ and $t_{eff}$. As the applied electric field increases, $n_{sheet}$ (blue) of the 2DEG layer monotonically increases and the $t_{eff}$ (green) becomes thinner. When $t_{eff}$ is thinner than the critical thickness, unusually large enhancement of $|S|$ is observed.

**Figure 2.** (a) Schematic illustration of the CAN-gated $SrTiO_3$ FET. Channel length and width are 800 and 400 μm, respectively. 200-nm-thick CAN film is used as the gate insulator. (b) Typical transfer characteristic curve of the CAN-gated $SrTiO_3$ FET at RT. A very large anticlockwise hysteresis of $I_d$ with a rather large $I_g$ was observed. (c, d) Cross-sectional TEM images of the CAN-gated $SrTiO_3$ FET at (c) before and (d) after applying positive $V_g$ (+40 V). Trilayer structure composed of Ti/CAN/$SrTiO_3$ is observed. Thickness of the CAN layer at the virgin state is ~200 nm, while it is ~550 nm, ~2.8 times thicker than the virgin state, after applying positive $V_g$. (e, f) Schematic 2DEG formation mechanism. For positive $V_g$, $H_3O^+$ ions in the CAN film strongly attracts residual electrons in the $SrTiO_3$ crystal, leading to the formation of 2DEG. After that, water electrolysis occurs at the $SrTiO_3$ and the Ti surfaces.

**Figure 3.** Changes in the thermopower and sheet carrier concentration for the 2DEG with gate voltage applications. Changes in (a) $|S|$ and (b) $n_{sheet}$ for the 2DEG layer as a function of $V_g$.

For comparison, changes in $|S|$ and $n_{sheet}$ for a dry C12A7-gated SrTiO$_3$-FET [11] as a function of $V_g$ at RT are also plotted. V-shaped turnaround of $|S|$ (indicated by the arrow) for the 2DEG layer is clearly observed. Retention time dependence of (c) $|S|$ and (d) $n_{sheet}$ for the 2DEG layer of the CAN-gated FET. $V_g$ is varied between +15, –15, –20, –30, and –40 V. For positive $V_g$ (+15 V), once $|S|$ decreases from 700 to 250 µV K$^{-1}$ with time (up to 500 s), it recovers drastically to ~600 µV K$^{-1}$. For negative $V_g$, $|S|$ gradually increases from ~600 to ~950 µV K$^{-1}$.

**Figure 4.** Electric field modulation of thermopower–sheet carrier concentration–effective thickness relation. $|S|$ values of the 2DEGs (Fig. 3a and 3c) are plotted as a function of (a) $n_{sheet}$ (Figs. 3b and 3d) and (b) $t_{eff}$ on a logarithmic scale. $|S|$ initially decreases with $n_{sheet}$ from ~1,150 to ~250 µV K$^{-1}$. Simultaneously, $t_{eff}$ decreases from ~100 to ~2 nm. In this region, the $n_{sheet}$ dependence of $|S|$ is similar to that of simulated bulk values (grey dotted line). When the $n_{sheet}$ value exceeds ~2.5×10$^{14}$ cm$^{-2}$, $|S|$ increases drastically and is modulated from ~600 to ~950 µV K$^{-1}$, while $t_{eff}$ remains nearly constant (~2 nm). $|S|$ vs. log $n_{sheet}$ relation is approximately five times larger than that of the bulk, clearly indicating that the electric field induced 2DEG in SrTiO$_3$ exhibits an unusually large $|S|$.

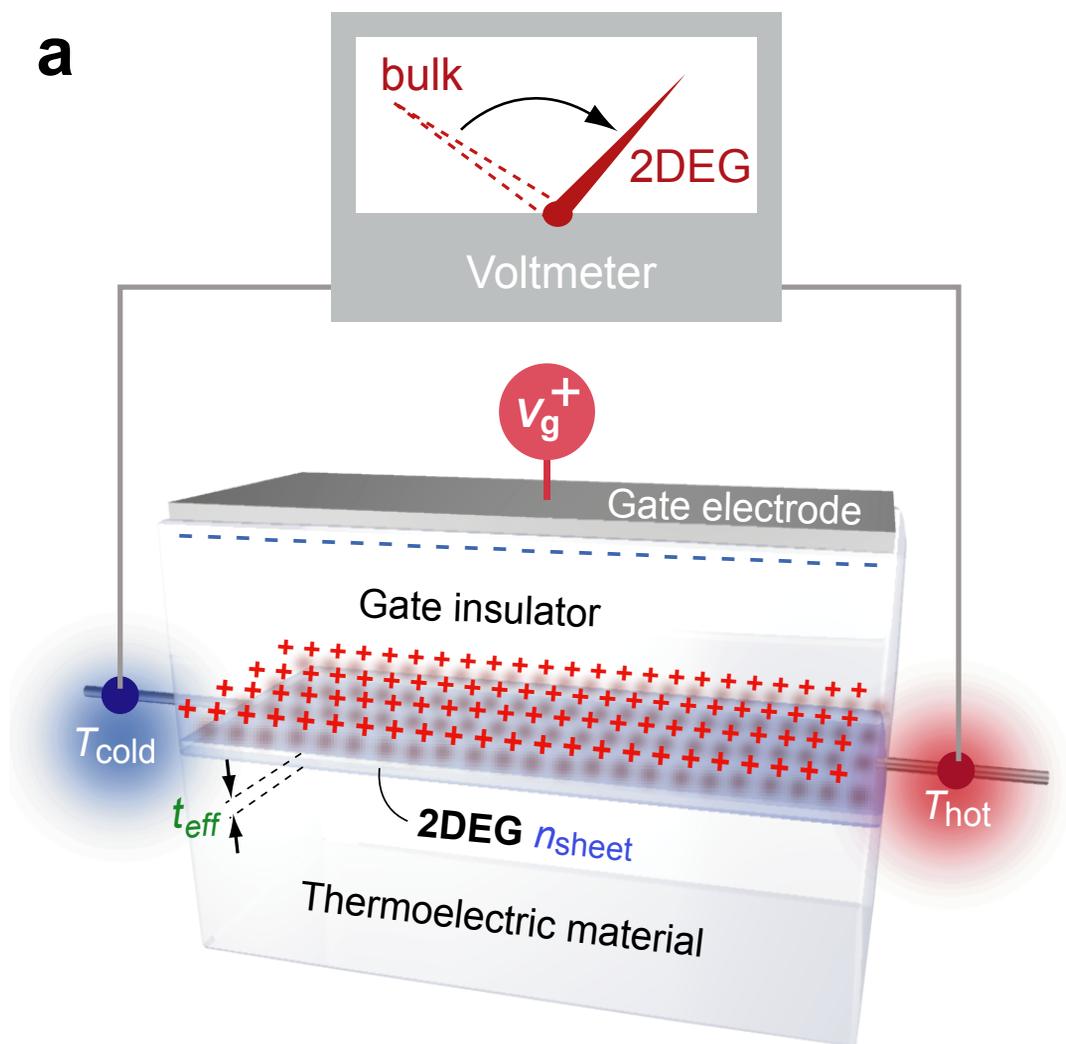
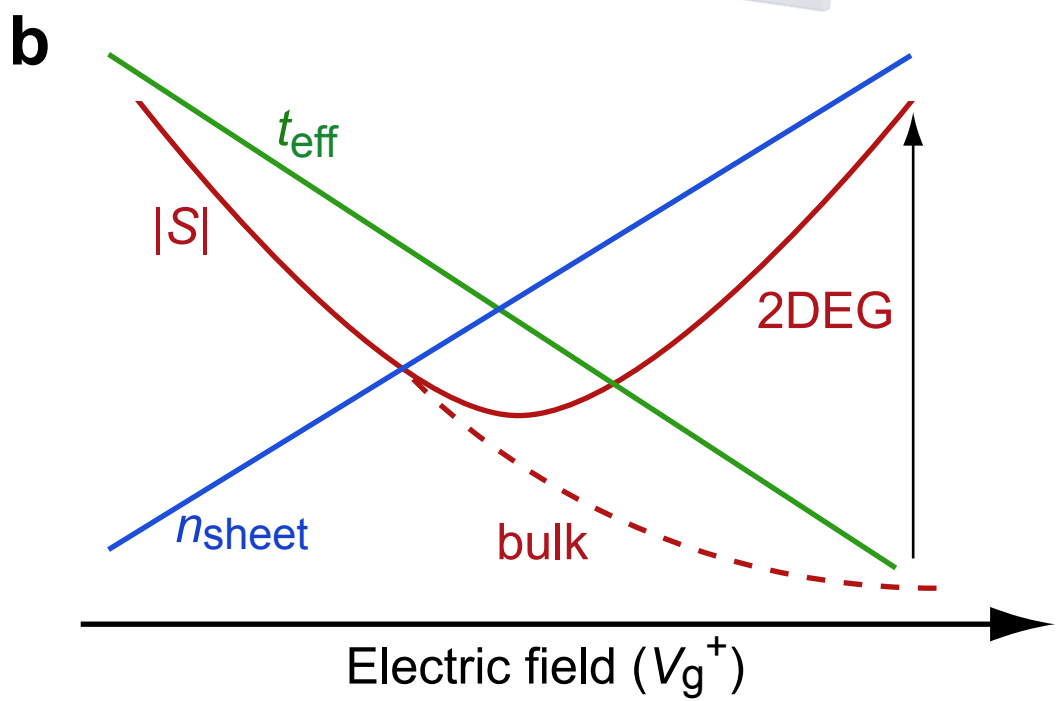

Fig. 1

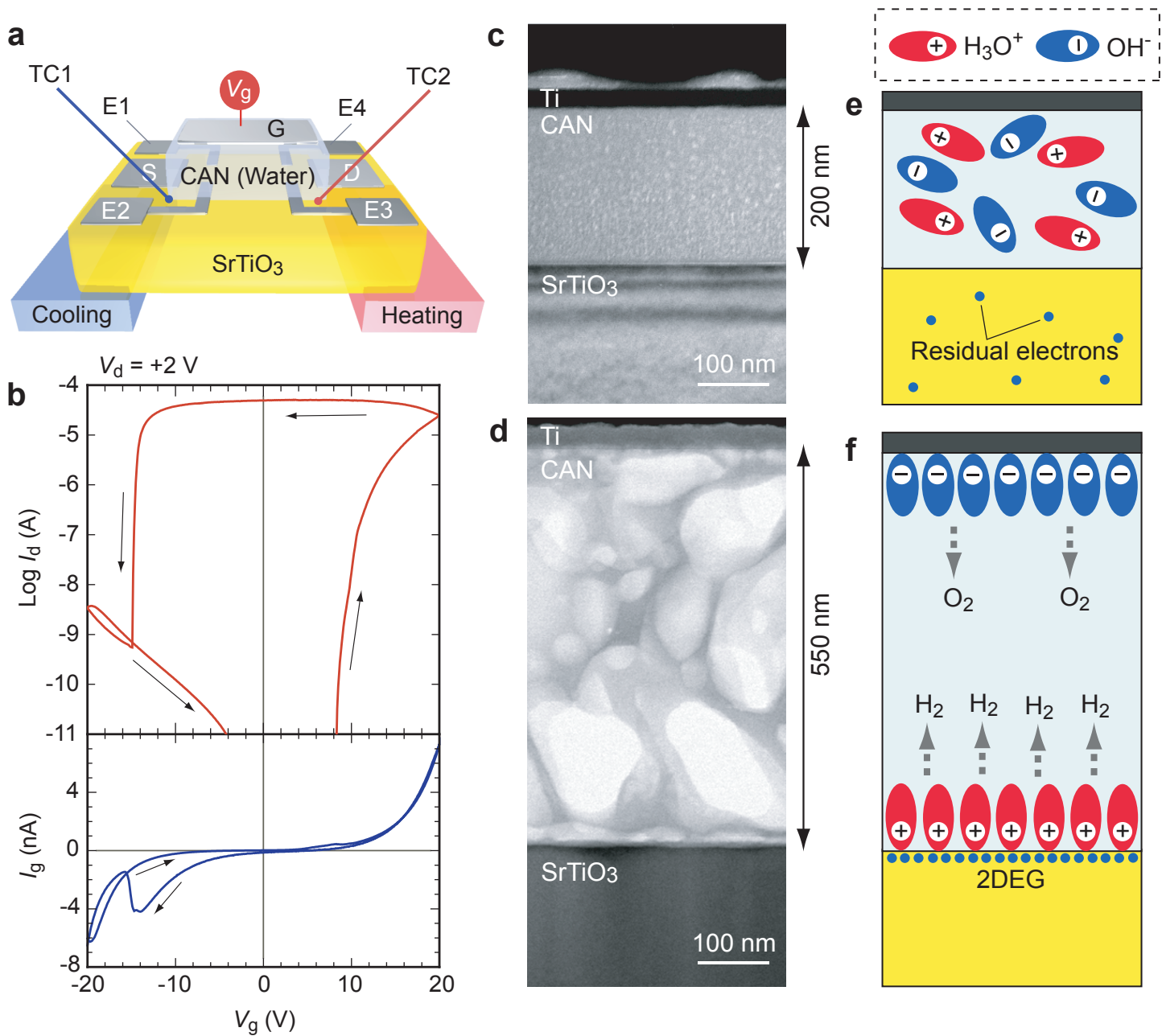

Fig. 2

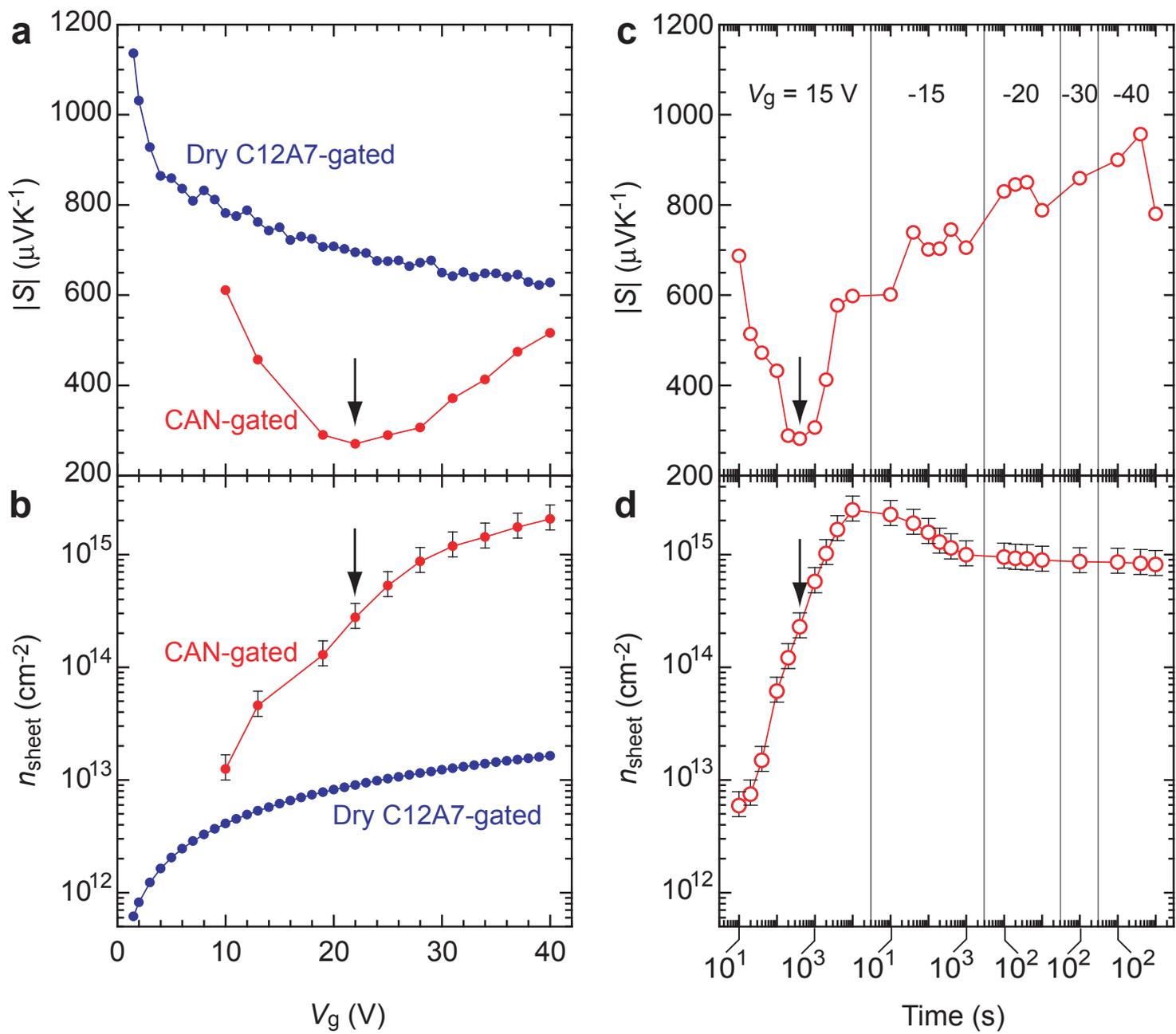

Fig. 3

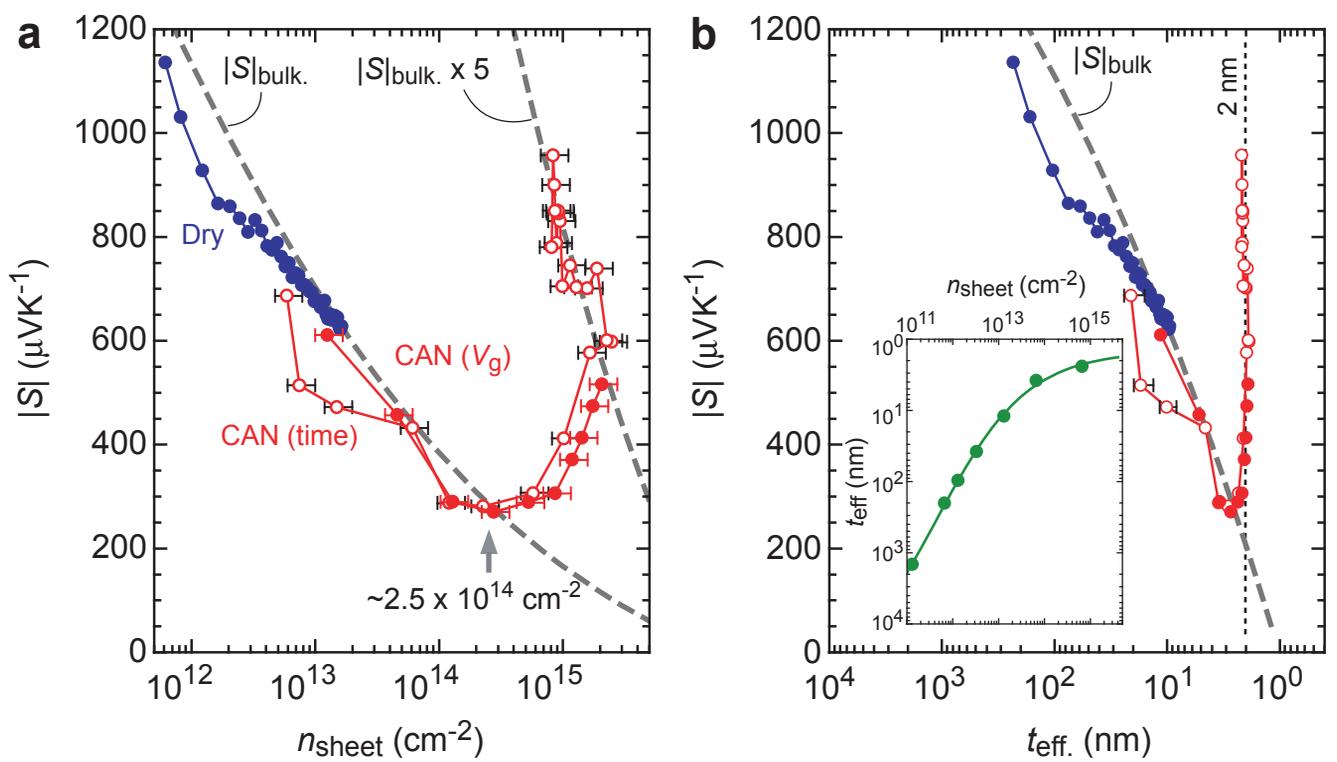

Fig. 4

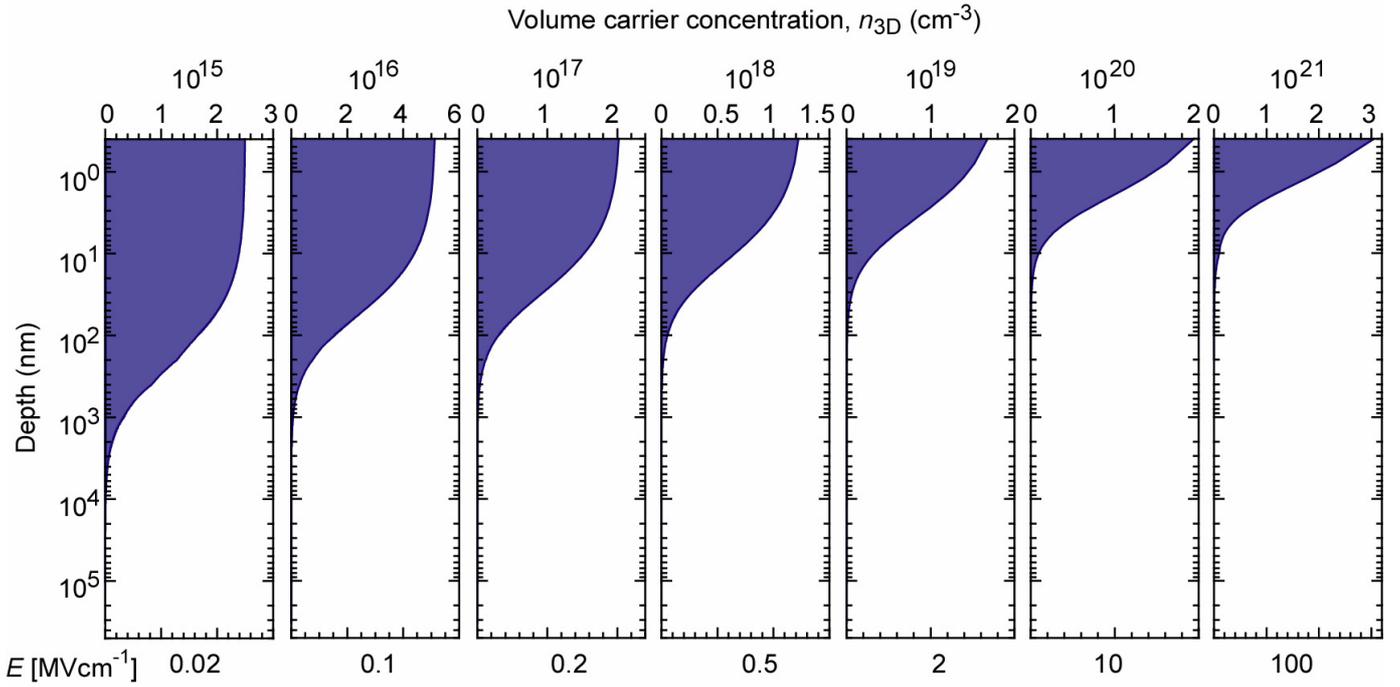

**Figure S1: Calculated carrier depth profile of Ti / *a*-C12A7 / SrTiO₃ FET.** The carrier depth profiles of the 2DEGs were calculated by using the commercially available ATLAS$^{TM}$, a 2D device simulator from SILVACO. Parameters used for the calculation is summarized in the following table.

|  | Ti | *a*-C12A7 | SrTiO₃ |
| --- | --- | --- | --- |
| Work function [eV] | 4.33 | ---- | 4.3 |
| Electron affinity [eV] | ---- | ---- | 4.1 |
| Bandgap [eV] | ---- | 6 | 3.3 |
| Residual carrier conc. [cm$^{-3}$] | ---- | ---- | $1 \times 10^{12}$ |
| Dielectric constant, $\varepsilon_r$ | ---- | 12 | 300 |

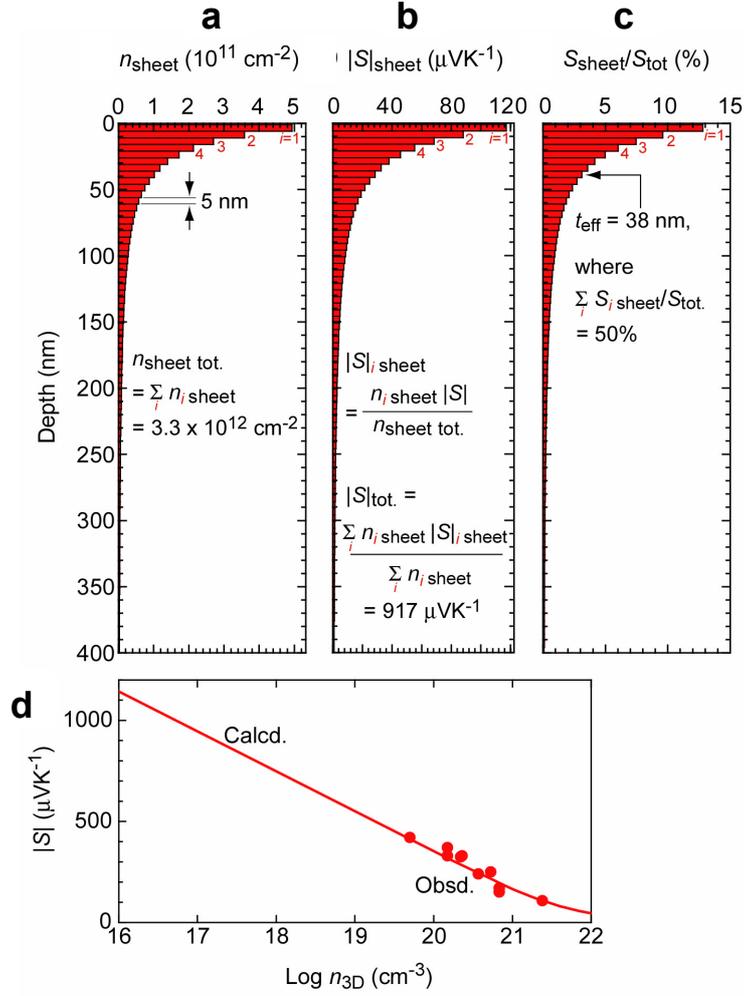

**Figure S2: Simulation of observable thermopower and effective thickness for the electric field induced 2DEG.** (**a**) Calculated carrier depth profile of the 2DEG ($E$=0.5 MVcm$^{-1}$). The 2DEG is expressed as multilayer of thin (5 nm, in this case) conducting layers with the 3D carrier concentration of $n_{i\ sheet}/5$ nm. The total sheet carrier concentration ($n_{sheet\ tot.}$) of the 2DEG can be expressed as

$$n_{sheet\ tot.} = \sum_i n_{i\ sheet}$$

, where $n_{i\ sheet}$ is the sheet carrier concentration of $i$-layer. (**b**) $|S|$-values of the $i$-layers ($|S|_{i\ sheet}$), which can be calculated using the $|S|$–Log $n_{3D}$ relation of bulk SrTiO$_3$ at RT (**d**) (Ref. 23) as

$$|S|_{i\ sheet} = \frac{\sigma_{i\ sheet}|S|}{\sigma_{sheet\ tot.}} = \frac{n_{i\ sheet}|S|}{n_{sheet\ tot.}}.$$

Since the $\mu_{Hall}$ values of the 2DEG were 2.4±0.6 cm$^2$V$^{-1}$s$^{-1}$, did not depend on the $n_{sheet}$, we assumed that the carrier mobility does not depend on the depth. Observable $|S|_{tot.}$ of the 2DEG can be calculated as $|S|_{tot.} = \dfrac{\sum_{i} n_{i\,sheet} |S|_{i\,sheet}}{\sum_{i} n_{i\,sheet}}$ . (**c**) $S_{sheet}/S_{tot.}$ We defined the effective thickness ($t_{eff.}$) of the 2DEG as the depth where $\sum_{i} S_{i\,sheet} / S_{tot.} = 50\%$.

**Table S1: Simulated relationship, electric field, sheet carrier concentration, thermopower and effective thickness.**

| $E$ (MVcm$^{-1}$) | $n_{sheet}$ (cm$^{-2}$) | $|S|$ (µVK$^{-1}$) | $t_{eff}$ (nm) |
|---|---|---|---|
| 0.02 | $1.3 \times 10^{11}$ | 1526 | 1466 |
| 0.1 | $6.6 \times 10^{11}$ | 1218 | 201 |
| 0.2 | $1.3 \times 10^{12}$ | 1080 | 96 |
| 0.5 | $3.3 \times 10^{12}$ | 917 | 38 |
| 2 | $1.3 \times 10^{13}$ | 682 | 12 |
| 10 | $6.6 \times 10^{13}$ | 426 | 3.8 |
| 100 | $6.6 \times 10^{14}$ | 206 | 2.4 |

**Reference**

23. H. Ohta, K. Sugiura, K. Koumoto, *Inorg. Chem.* **2008**, 47, 8429.